\documentclass[%
 reprint,
superscriptaddress,
 amsmath,amssymb,
 aps,
floatfix,
]{revtex4-2}

\usepackage{graphicx}
\usepackage{dcolumn}
\usepackage{bm}
\usepackage{hyperref}
\hypersetup{
    colorlinks=true,
    linkcolor=blue,
    citecolor=blue,
    filecolor=magenta,      
    urlcolor=black,
    pdftitle={Overleaf Example},
    pdfpagemode=FullScreen,
    }
\usepackage{tikz}

\usepackage[mathscr]{euscript}
\begin{document}

\preprint{APS/123-QED}

\title{Spreading of a 2D granular analogue of a liquid puddle: \\ predicting the structure and dynamics through a continuum model}

\author{Johnathan Hoggarth}
\affiliation{\mbox{Department of Physics and Astronomy, McMaster University, 1280 Main Street West, Hamilton, Ontario, L8S 4M1, Canada}}
\author{Jean-Christophe Ono-dit-biot}
\affiliation{\mbox{Department of Physics and Astronomy, McMaster University, 1280 Main Street West, Hamilton, Ontario, L8S 4M1, Canada}}
\author{Kari Dalnoki-Veress}
\email{dalnoki@mcmaster.ca}
\affiliation{\mbox{Department of Physics and Astronomy, McMaster University, 1280 Main Street West, Hamilton, Ontario, L8S 4M1, Canada}}
\affiliation{UMR CNRS Gulliver 7083, ESPCI Paris, PSL Research University, 75005 Paris, France}

\date{\today}

\begin{abstract}
When sand flows out of a funnel onto a surface, a three dimensional pile that is stabilized by friction grows taller as it spreads. Here we investigate an idealized two dimensional analogue:  spreading of a pile of monodisperse oil droplets at a boundary. In our system the droplets are buoyant, adhesive, and in contrast to sand, here friction is negligible. The buoyant droplets are added to the pile one-at-a-time. As the aggregate grows, it reaches a critical height and the 2D pile spreads out across the barrier. We find that, while granularity is important, the growth process is reminiscent of a continuum liquid. We define a ``granular capillary length'', analogous to the capillary length in liquids, which sets the critical height of the aggregate through a balance of buoyancy and adhesion. At a coarse-grained level, the granular capillary length is capable of describing both steady-state characteristics and dynamic properties of the system, while at a granular level repeated collapsing events play a critical role in the formation of the pile.
\end{abstract}

\maketitle


\section{Introduction}
 
From making sandcastles at the beach to storing grains in large silos, piles of granular materials are a familiar sight which can span vast length scales. Granular materials are a particularly interesting class of materials as they can exhibit features of both solids and liquids, depending on the forces acting on the material, with rich phenomena. These systems exhibit unique properties such as the presence of characteristic angles of repose~\cite{BrunoAndreottiYoelForterre2013, Jaeger1996}, intermittent flow resulting in avalanches~\cite{GDR, Lemieux2000, Frette1996}, clogging and jamming~\cite{Liu1995, Liu1998, Cates1998, Zuriguel2014, Dressaire2017,Behringer_2018}, and the presence of force chains~\cite{Mueth1998, Vallejo2005}.

The main focus of this work is closely related to the familiar case of pouring sand out of  a funnel onto a surface. As the sand hits the barrier, a pile grows taller and spreads. There have been many studies examining how a granular material accumulates at a barrier and how the angle of repose is impacted when material properties change ~\cite{BeakawiAl-Hashemi2018,Lanzerstorfer2017, Sarate2022, Fu2020}. For granular materials consisting of dry cohesionless grains, the angle of repose is primarily dependent on the internal friction of a material and it also has a dependence on particle size, shape, and roughness~\cite{Fu2020, Zhou, Combarros2014, Hsiao2019}. Investigations into more complex systems where particles are influenced by additional forces such as cohesion~\cite{Hornbaker1997,Gans2020, Halsey1998, Nowak2005, Richefeu2006}, thermal motion~\cite{Naima2016, Berut2019}, and particle flow~\cite{RobertDeSaintVincent2016, Sanchez2007, Alvaro2011}, show that the structure of accumulations of granular materials are robust.

Of particular interest are studies showing that even in systems where friction is negligible, spheres can accumulate with a measurable angle of repose~\cite{Ortiz2013, Shorts2018, Lespiat}. Ortiz \textit{et al.} observed that slightly repulsive frictionless colloids would accumulate with a measurable angle of repose while accumulating at a barrier in a microfluidic channel~\cite{Ortiz2013}. Shorts and Feitosa were able to measure an angle of repose as air bubbles accumulated at the surface of a chamber~\cite{Shorts2018}. Results in both of these experiments showed that particle flow -- a continuous replenishing of particles -- was required to observe the presence of an angle of repose. Lespiat \textit{et al.} showed that the properties that exist in frictionless foams are governed by analogous laws that exist in hard spheres, including the presence of a critical angle where flow begins to occur~\cite{Lespiat}. Additionally, the impact of friction on the critical angle required to cause motion in a granular material has been investigated both experimentally~\cite{Perrin2019, Perrin2021} and numerically~\cite{Peyneau2008} showing that, while different from a frictional case, this critical angle is still present.

In contrast to the growth of a granular pile which gets taller as it spread, a liquid drop will grow at a boundary in two stages: In the first stage, as liquid is added a droplet will grow as a spherical cap due to surface tension. As the droplet gets larger, it will reach a critical height, due to gravity, which is determined through a balance of the effects of surface tension with gravity~\cite{DeGennes2004}. The characteristic scale is set by the capillary length and  defined as $\kappa^{-1}=\sqrt{\gamma/\rho g}$ where $\gamma$ is the interfacial tension, $\rho$ is the density of the liquid, and $g$ is the gravitational acceleration. 

Here we describe an experiment observing the accumulation of monodisperse oil droplets at a barrier in two-dimensions. The oil droplets are buoyant and friction is negligible. Additionally, the oil droplets experience a tuneable attractive force due to the depletion interaction resulting from micelles present in the aqueous phase. The oil droplets are produced one at a time and, due to buoyancy, will rise to the top of a chamber which consists of a flat glass slide held at some fixed angle. The droplets then move along the glass plane along the direction of steepest ascent until they are stopped by a barrier where they form an aggregate. With the experiments performed we have the ability to tune both the effective buoyant and adhesive forces acting on the droplets by adjusting the angle of the chamber and the micelle concentration. This experiment builds upon our previous work where we investigated the accumulation of an aggregate of oil droplets in three-dimensions~\cite{Ono-dit-Biot2020}. In that study we found that the spreading of the aggregate at a flat horizontal interface exhibited properties similar to that of the growth of a liquid puddle. We found that the equilibrium shape of an aggregate of oil droplets could be described through a parameter which was termed the \emph{granular capillary length}, $\delta$. By analogy with the capillary length of a liquid, the granular capillary length is determined through a balance between a droplet's effective buoyancy and the adhesive strength between the droplets.

In the current paper, we see that the granular capillary length can be extended as a fundamental parameter for a two-dimensional system and performs well as a predictor for both the equilibrium structure and the dynamic properties of aggregates of oil droplets as they spread in two-dimensions.

\section{Experimental Setup}
\begin{figure}
    \includegraphics[width=1\columnwidth]{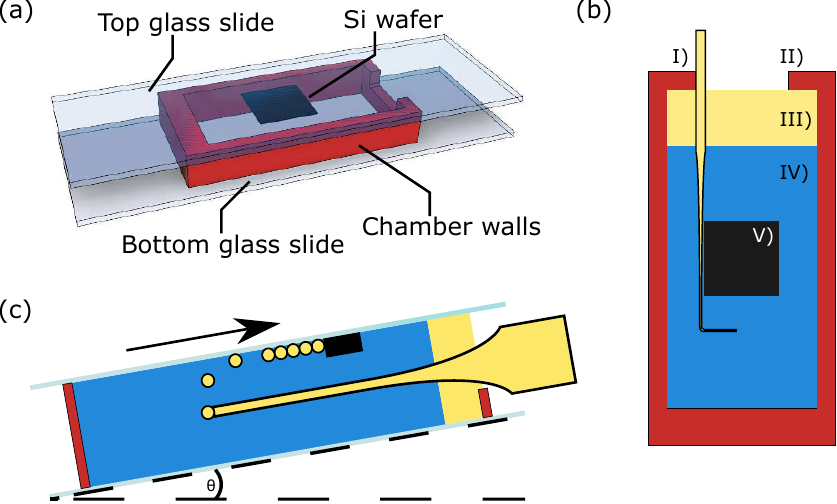}
    \caption{\label{fig:Samplechamber} (a) A schematic diagram of the experimental chamber. All of the materials of the sample chamber are held together using silicone caulking. (b) A top-down schematic of a chamber I) droplet producing micropipette, II) chamber walls, III) a layer of mineral oil to prevent evaporation, IV) aqueous solution of SDS, and V) cleaved Si wafer. (c) A schematic diagram of an experiment in progress. The dashed line shows the angle $\theta$ of the chamber which impacts the effective force of buoyancy. The arrow shows the direction of the effective force of buoyancy acting on the aggregate.}
\end{figure}

The experimental setup, as illustrated in FIG.~\ref{fig:Samplechamber}, consists of chamber, made of two glass slides separated by a 3D printed spacer with an inner volume of 16 x 34 x 6 mm$^3$. A 100 $\mu$m thick wafer of silicon (Si) (University Wafer, USA) was cleaved into approximately 5 x 5 mm$^2$ wafers and attached to the top glass slide with silicone caulking to act as a barrier. A simple microscope (camera and objective) is mounted above the chamber and normal to the plane of the glass slide. The chamber was partially filled with an aqueous solution of sodium dodecyl sulfate (SDS) and NaCl. The concentration of SDS, $C$ varied across experiments from 1\% to 3\% (w/w) ($\sim 20$~mM to $100$~mM). The concentration of NaCl was held constant across all experiments at 1.5\% (w/w). At these concentrations of SDS, not only will the SDS stabilize the oil droplets as a surfactant, but it will also form micelles above the critical micelle concentration, $C_{\mathrm{CMC}}$,  which induce attractive interactions between oil droplets due to the depletion interaction~\cite{Bibette1992,jones2002soft}. The concentration of micelles can be calculated, $C_{\mathrm{m}} = C - C_{\mathrm{CMC}}$, where $C_{\mathrm{CMC}}$= 8 mM in pure water ~\cite{Bibette1992,Thevenot2005, Naskar2013}. While the concentration of micelles sets the attractive interaction between the oil droplets, the NaCl is added to the solution to screen ionic interactions caused by the polarity of SDS. While it is known that the addition of NaCl to a solution of pure water slightly decreases the CMC of SDS, this change is negligible for our calculations ~\cite{Thevenot2005,Naskar2013}. Mineral oil was also added to the chamber to prevent evaporation of the SDS solution (FIG.~\ref{fig:Samplechamber}(b)). Chambers were reused several times over the course of several days of experimentation without affect on the results.

Droplets are prepared using the snap-off instability which results in monodisperse droplets~\cite{Barkley2015,Barkley2016} as follows. Glass capillary tubes with outer diameter 1~mm and inner diameter of 0.58~mm (World Precision Instruments, USA) were pulled using a pipette puller (Narishige, Japan) to create a micropipette. The tip of the pipette was broken to produce an opening with a diameter of $\sim 10 \ \mu$m. The size of the opening could be adjusted by breaking the micropipette tip at different places. The pipette is inserted into a gap in the chamber (FIG.~\ref{fig:Samplechamber}) and the other end is connected to a syringe containing mineral oil. Mineral oil is pushed through the pipette to create the monodisperse oil droplets with sizes proportional to the pipette tip diameter~\cite{Barkley2015,Barkley2016}. Once the oil droplets begin to emerge with the high pressure provided by the syringe, the syringe is removed. The plastic housing to which the syringe was attached forms a small reservoir of mineral oil open to atmospheric pressure, the height of which can be adjusted to control the production rate of droplets. The droplets are buoyant so they rise to the top of the chamber and will move along the top glass slide until they contact the Si barrier where they will accumulate. For all experiments, the height of the reservoir was positioned to obtain a rate of roughly one droplet every 30 seconds. This slow rate ensures that the aggregate has enough time to reach a state of quasi-equilibrium prior to the addition of another droplet. The chamber and microscope are attached to a motorized rotation stage such that the angle of the chamber, and thus the effective buoyant force acting on the droplets, can be controlled while keeping the camera normal to the surface. 

\begin{figure}
    \includegraphics[width=1\columnwidth]{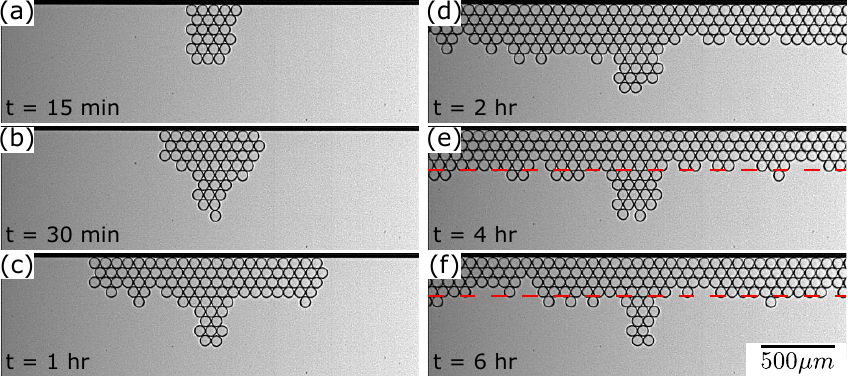}
    \caption{\label{fig:evolution} (a)-(f) Accumulation of an aggregate of oil droplets over time at a boundary. $R$ = 37 $\mu$m, $\theta$ = 30$^{\circ}$, $C_{\mathrm{m}}$~= 63 mM. Time elapsed for each image is shown at the bottom left. The red dashed line in panels (e) and (f) show the steady state height of the droplet pile.}
\end{figure}

\section{Results and Discussion}

\subsection{Spreading of Droplet Aggregates}

\begin{figure}
    \includegraphics[width=1\columnwidth]{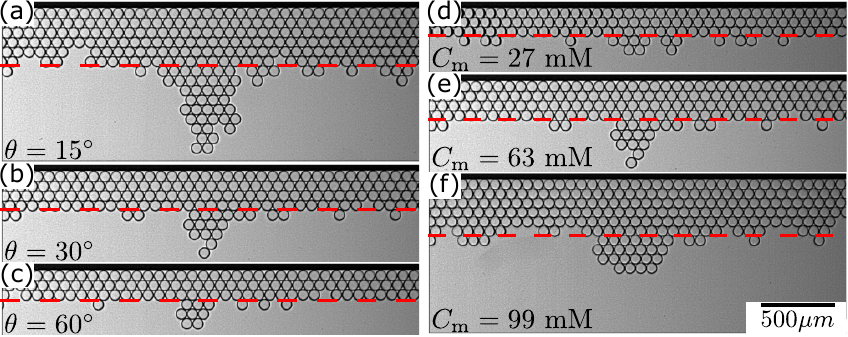}
    \caption{\label{fig:theta_c} Left side: three experiment snapshots showing the aggregate with the top plate of the chamber tilted at (a-c) $\theta=15^\circ$,  $30^\circ$, and $60^\circ$, with  $C_{\mathrm{m}}= 63  $~mM and $R=37 \  \mu$m. Right side: three experiment snapshots showing the aggregate with changing SDS concentration (d-f) $C_{\mathrm{m}}= 27$~mM, 63~mM, and 99~mM, with  $R = 37 \ \mu$m and $\theta =30^\circ$.}
\end{figure}

The evolution of the 2D aggregate at a boundary is shown in FIG.~\ref{fig:evolution}. The aggregate neither forms a typical conical pile, as expected for a granular material, nor spreads out continuously across the barrier. Instead, the aggregate can be described by two distinct regions: First, there is the central region where droplets are added which grows until it reaches a critical height at which point adhesive bonds between the droplets are broken and this region collapses. Beyond that, on either side of the aggregate there is the bulk of the 2D crystal which spreads out across the Si barrier with a constant average height. We note that the barrier is of a finite length and about two orders of magnitude longer than the droplet diameter. When the droplets reach the edge of the barrier they simply detach off the sides and do not affect any of the physics of interest here. We first focus on the bulk of the aggregate, ignoring the central region which is described by repeated collapsing events due to the granular nature of the system. The bulk region is in a state of quasi-equilibrium, as its structure remains constant in height. The process of spreading for the bulk of the aggregate can be seen as analogous to a continuous liquid spreading, since it reaches a steady-state height and spreads out across the barrier. The steady-state  height was calculated over a 2 hour period once the aggregate had spread along the entire barrier and stabilized (see snapshots in FIG.~\ref{fig:evolution}(d)-(f)).
In the experiments the angle of the chamber, $\theta$, the radius of the droplets, $R$, and the concentration of SDS micelles, $C_{\mathrm{m}}$, were varied to examine the dependence of these parameters on the aggregate geometry. We assume that the friction and viscous dissipation is negligible and therefore changing these three parameters only act to change the forces of adhesion and buoyancy acting on the droplets.  As will be seen, and consistent with the work by Ono-dit-Biot~\cite{Ono-dit-Biot2020,Ono-dit-Biot2021PRR,Ono-dit-Biot2021SM}, these assumptions are valid: the droplets are added to the structure slowly, thus dissipation due to viscosity does not play a role; furthermore, there is no friction at the fluid interfaces. As the angle of the chamber is increased, and therefore the effective buoyancy ($\propto \sin\theta)$ acting on the pile is increased, the steady-state height of the aggregate decreased as shown in FIG.~\ref{fig:theta_c} (a)-(c). The observations are sensible since any increase in the height of a pile must act against the force of buoyancy. Additionally, for a given angle, it was seen that increasing the concentration of SDS, therefore increasing $C_{\mathrm{m}}$ and thus the adhesive forces between droplets increased the steady-state height as shown in FIG.~\ref{fig:theta_c} (d)-(f).

\subsection{Aggregate density profile}

\begin{figure}[b]
	\centering
	\includegraphics[width=1\columnwidth]{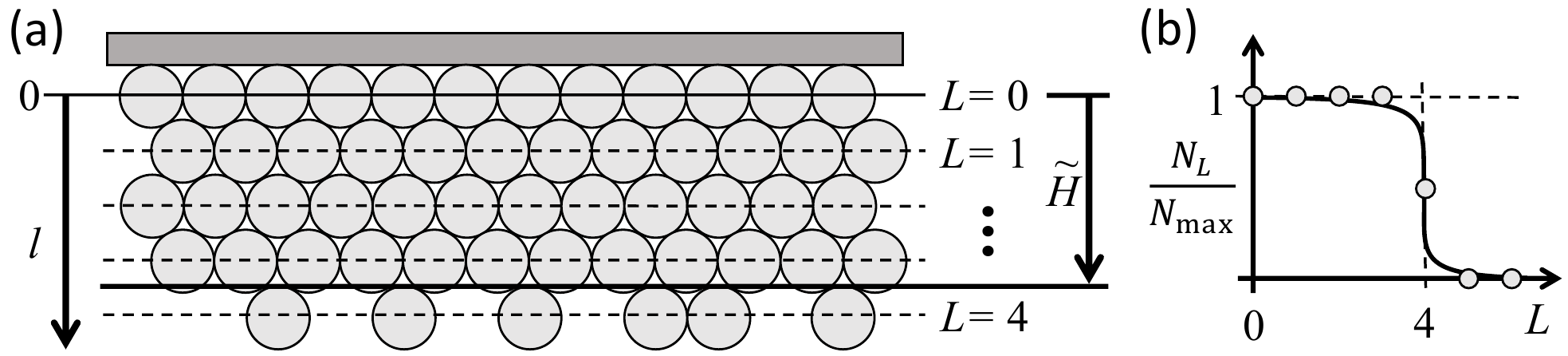}
    	\caption{\label{fig:schem} (a) Schematic of an aggregate with the symbols used in the text. Note that $l$ is a length, while $L$ is an index associated with the layer number, and $\widetilde{H}$ is the non-dimensional height of the structure representing the number of layers beyond the first $L=0$. (b) Schematic of the layer density profile as a function of the layer number.} 
\end{figure}

Having seen the qualitative dependence of the aggregate height upon independently changing the adhesion strength and the effective buoyancy, we now focus on understanding the profile of the density of the droplets as a function of the distance from the barrier in greater detail. We define the steady-state non-dimensional height $\widetilde{H}$ as shown in Fig~\ref{fig:schem}, which represents the non-dimensional average number of layers in an aggregate beyond the first layer $L=0$. We have $\widetilde{H}=\sum_{L=1}^{\infty}{N_L}/{N_{\mathrm{max}}}$, where $N_L$ is the number of droplets in layer $L$, and $N_{\mathrm{max}}$ is the maximum number of droplets that can occupy a layer.   In the schematic shown in Fig~\ref{fig:schem}(a), $\widetilde{H}=1+1+1+0.5=3.5$ since layer $L=4$ is half occupied. The non-dimensional variables can be related to the dimensional distance from the midpoint of the $L=0$ layer as $l=L\sqrt{3}R$ and the height $h=\widetilde{H}\sqrt{3}R$.

In the experiments there are random fluctuations of the aggregate height. This can be seen in  FIG. \ref{fig:evolution}(e) and (f)  where there were droplets above and below the steady-state height indicated by the dashed line. We define these fluctuations as the non-dimensional width $\widetilde{w}$ of the transition from the layers with a maximum ``layer density'' ${N_L}/{N_{\mathrm{max}}}=1$ to the region where there are no droplets ${N_L}/{N_{\mathrm{max}}}=0$ (see FIG.~\ref{fig:schem}(b)). This transition region is analogous to the interfacial width of transition from the condensed phase to the vapour phase in molecular systems. The width is obtained from the data by assuming that the interfacial profile of the aggregate follows a hyperbolic tangent profile. The layer density profile as a function of the layer number is then given by the expression:
\begin{equation}
    \frac{N_L}{N_{\mathrm{max}}} =\frac{1}{2}\left[ 1- \tanh{\left(\frac{\widetilde{H} + 1/2-L}{\widetilde{w}}\right)} \right]
    \label{eqn:tanh_profile}
\end{equation} 
(note that the additive constant $1/2$ in the hyperbolic tangent term arises from the fact that since $\widetilde{H}=3.5$, layer $L=4$ is half filled as in the example in Fig~\ref{fig:schem}).

\begin{figure}
    \includegraphics[width=1\columnwidth]{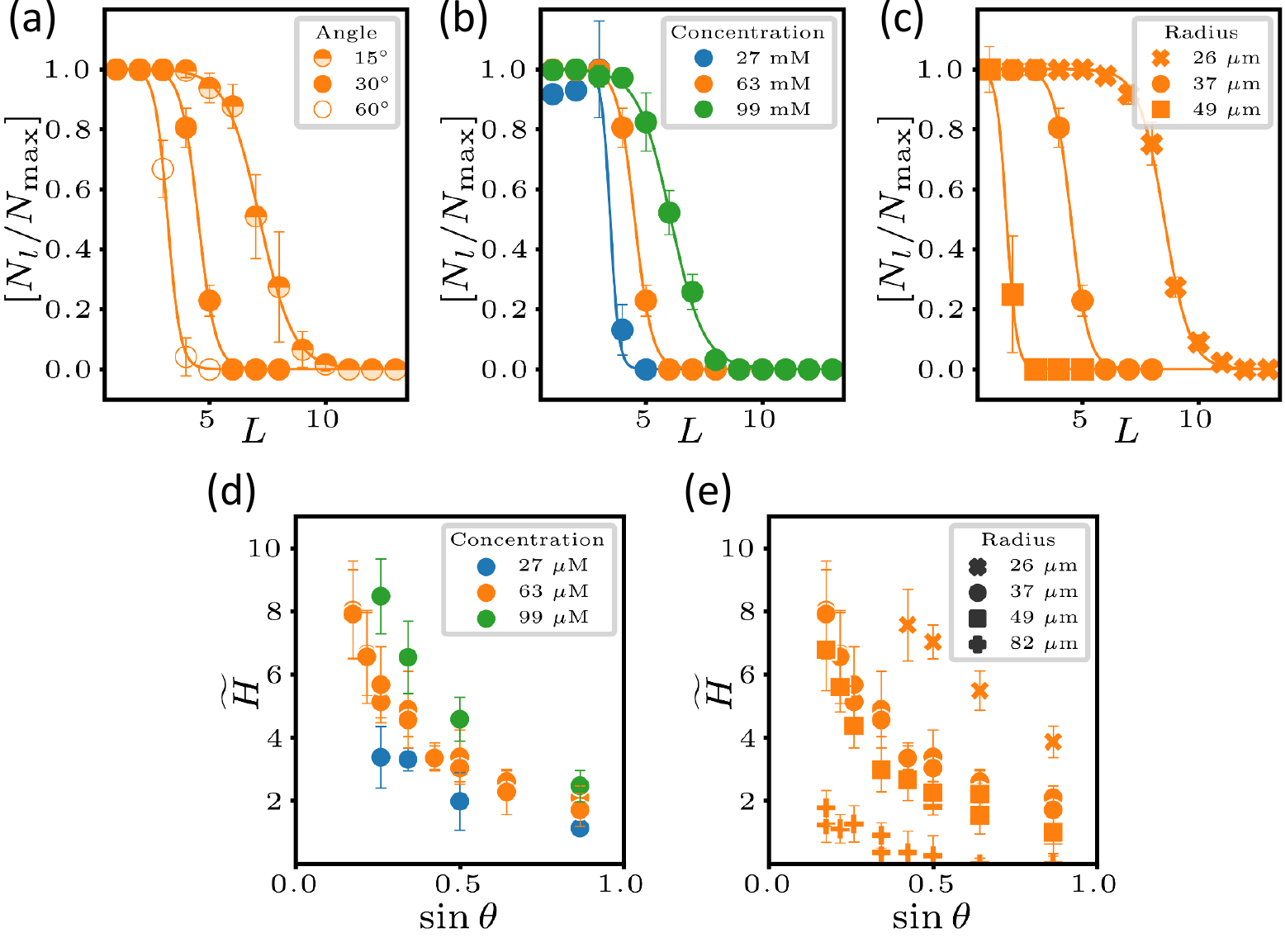}
    \caption{\label{fig:H_w}
(a)-(c) Normalized average number of droplets in a layer plotted as a function of the layer number. The solid lines are a best fit of Equation \ref{eqn:tanh_profile} to the data. Error bars are the standard deviation in the averaged data varying the angle (a), the concentration (b) and the radius (c).
Average steady-state height of the pile plotted as a function of the effective buoyancy force $ \propto \sin{\theta}$ for different concentrations of $C_{\mathrm{m}}$ with constant radius $R = 37 \ \mu$m  (d); and different droplet radii with a constant concentration of $C_{\mathrm{m}} = 63 \ $mM (e).} 
\end{figure}

When plotting the layer density, ${N_L}/{N_{\mathrm{max}}}$, averaged over $2$ hours after the pile has stabilized to a steady-state, as a function of the layer number $L$, we find that the interfacial width is well described by equation \ref{eqn:tanh_profile} as shown in FIG.~\ref{fig:H_w}(a)-(c). For each experiment the average value of the non-dimensional width, $\widetilde{w}$ and height $\widetilde{H}$, was obtained by fitting  equation \ref{eqn:tanh_profile} to the averaged data. From Fig.~\ref{fig:H_w} several points are apparent: i) Upon decreasing the angle, so the chamber is more horizontal, the effective buoyancy decreases and as a result both $\widetilde{H}$ and $\widetilde{w}$ increase (see Fig.~\ref{fig:H_w} (a)). ii) Upon increasing the adhesion, the aggregate is more stable and $\widetilde{H}$ and $\widetilde{w}$ increase (see Fig.~\ref{fig:H_w} (b)). iii) And, lastly, increasing the size of the oil droplets increases both the effective force of buoyancy and the adhesive forces between the droplets (see Fig.~\ref{fig:H_w} (c)). However, since the effective force due to buoyancy scales with $R^3$, while the adhesion strength scales as $R$ ~\cite{Ono-dit-Biot2020},  upon increasing the droplet size both $\widetilde{H}$ and $\widetilde{w}$ decrease.

\subsection{Granular Capillary Length}

\begin{figure}
    \includegraphics{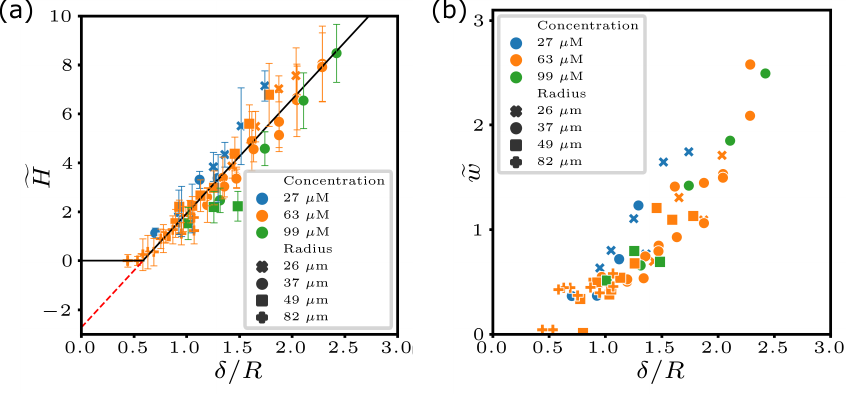}
    \caption{\label{fig:deltaPlots} (a) Plot of $\widetilde{H}$ as a function of $\delta/R$ for all experiments. The solid line represents Equation \ref{eqn:height}. (b) Plot of the width of the interface as a function of $\delta/R$ for all experiments. Error bars for each plot represent the standard deviation in the data. }
\end{figure}

We have shown that the structure of the bulk of the aggregate depends on a balance of buoyancy and adhesion. Previous work has shown that we can extend the concept of a capillary length, which arises from a balance between surface tension and gravity, to a granular capillary length~\cite{Ono-dit-Biot2020}. This length scale arises naturally as a balance between the effect of gravity, through the effective buoyancy of the droplets, and the adhesion between the droplets:
\begin{equation}
    \delta = \sqrt{\frac{\mathscr{A}}{\Delta\rho g \sin \theta}},
    \label{eqn:delta}
\end{equation} 
where $\mathscr{A}$ is the adhesion between droplets, $\Delta\rho$ is the difference in densities between the oil and SDS, and $g$ is the acceleration due to gravity. The exact value for the adhesion strength were previously measured for oil droplets of various sizes and concentrations of SDS~\cite{Ono-dit-Biot2020}. 

Extending our analogy to continuum liquids, we assume that the maximum height that can be sustained by the bulk of the aggregate of oil droplets is proportional to the granular capillary length. Stability of a layer that is a distance $h+R$ from the barrier, requires that the addition of one more layer which increases the height by $\sqrt{3}R$ must then be unstable. We note that in the continuum limit where $R\rightarrow0$, the stability criterion is $h = \alpha \delta$ as expected. Thus the stability criterion for the discretized system is that $h+R+\sqrt{3}R < \alpha \delta$, where $\alpha$ is a proportionality constant of order one. Since $h=\widetilde{H}\sqrt{3}R$ we can then express the boundary of stability simply as:
\begin{equation}
   \widetilde{H} = \frac{\alpha}{\sqrt{3}} \cdot \frac{\delta}{R} - \left(1+\frac{1}{\sqrt{3}}\right).
    \label{eqn:height}
\end{equation}
A plot of $\widetilde{H}$ as a function of $\delta/R$ is shown in FIG.~\ref{fig:deltaPlots}(a) for 67 experiments with 3 different concentrations, 4 different droplet radii, and 9 different angles of the chamber. We see an excellent collapse of all the data which indicates that the height of the pile is well explained by the granular capillary length. The solid line is a fit of Equation \ref{eqn:height} with only one free parameter, the proportionality constant $\alpha = 4.66 \pm 0.06$. We note that as expected $\alpha$ is of order one. Furthermore, it is interesting to note that for values of $\delta/R \le (1+\sqrt{3})/\alpha$, $ \widetilde{H}=0$ since even with no adhesion one row of droplets is stable. Lastly, given that the width of the interface should depend on buoyancy and adhesion, we plot the width as a function of the characteristic length-scale of the system given by the granular capillary length in FIG. \ref{fig:deltaPlots}(b). The data collapses which is consistent with the suggestion that the width can be captured through the granular capillary length. The width is seen to vanishes for finite values of  the granular capillary length in the data since only a single row is stable for $\delta/R \le (1+\sqrt{3})/\alpha$.

\subsection{Dynamics of Spreading}

Following our observations of the steady-state height, $\widetilde{H}$, and width, $\widetilde{w}$,  the spreading of the aggregates relies on sudden fractures and collapsing events. Once the central pile reaches a critical height, a collapsing event occurs much like how an avalanche will occur in granular system above a critical angle. 

\begin{figure}
    \includegraphics[width=1\columnwidth]{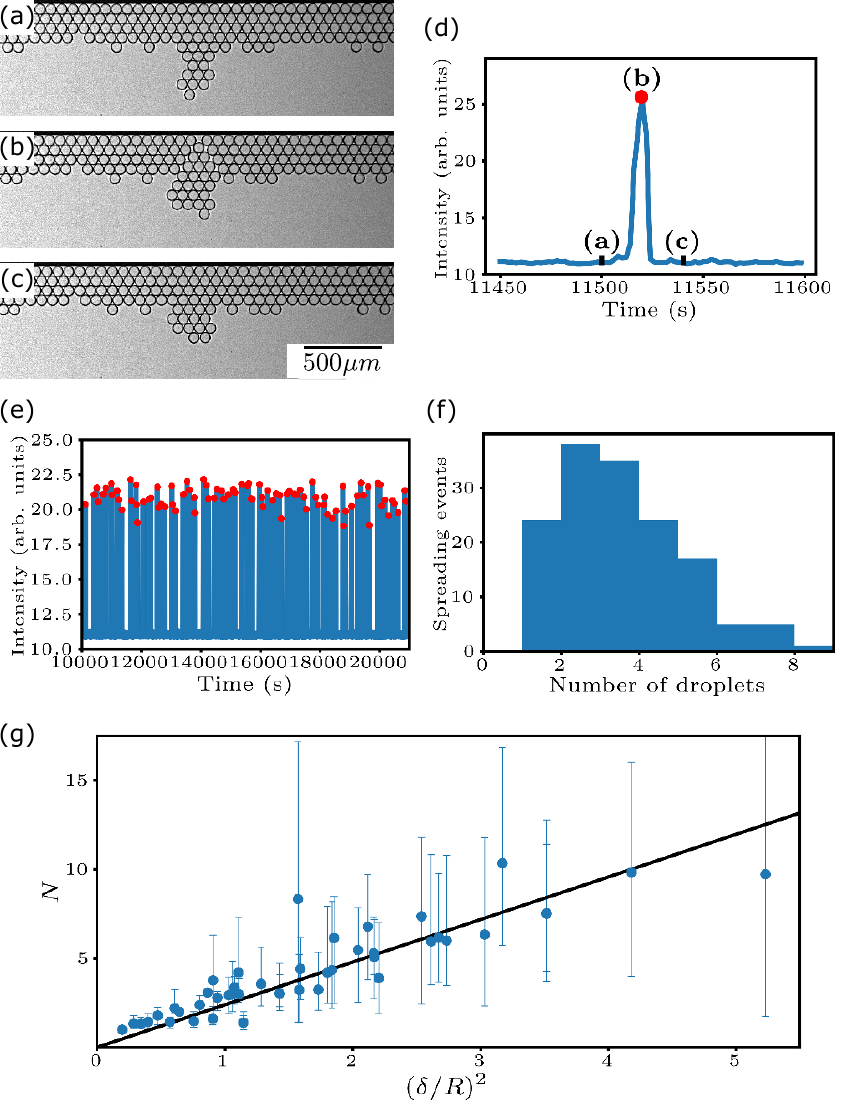}
    \caption{\label{fig:avalanches} (a)-(c) Images of an aggregate undergoing a spreading event. (d) Average intensity of subtracted images with red dot showing the location of a singe spreading event. (e) Average intensity of subtracted images of the entire time span after the aggregate has stabilized. (f) Distribution of the number of droplets required to cause a spreading event. (g) Median number of droplets required to cause a spreading event plotted against $(\delta/R)^2$. Experiments with less than 15 avalanches were excluded. Error bars represent first and third quartiles.}
\end{figure}

We find that while the bulk of the material spreads at a steady-state height, the central region grows to a critical point at which point bonds between droplets are broken and the bulk of the aggregate spreads out. We define a spreading event as any event caused by the addition of a droplet to the central region which leads to further spreading of the pile. To determine when such an event occurs, we observe the motion of the pile over time. Each image of the time series of our experiment is subtracted from the previous image such that the average intensity of these subtracted images is a proxy for droplet motion. The average intensity for a single spreading event is plotted over time and it can be seen that a peak in this intensity corresponds to a spreading event occurring in the aggregate as shown in FIG.~ \ref{fig:avalanches}(a) - (d). As droplets are added to the pile, multiple spreading events occur and the distribution of the number of droplets required to cause a spreading event can be obtained as shown in FIG. ~\ref{fig:avalanches} (e) and (f).

To explain the dynamics of the spreading, the size of the central pile of the aggregate is observed over time. As more droplets are added, the area of the central pile, $A$, will grow proportionally to the number of droplets added $N$ such that $A \propto N$. Since we assert that the granular capillary length is a fundamental length scale of our system, we also assume that the area of the central pile must scale as the square of the granular capillary length: $A \propto (\delta/R)^2$. Therefore the number of droplets added to the central region before a spreading event must be set by the granular capillary length, 
\begin{equation}
    N \propto (\delta/R)^2.
    \label{eqn:avalanche}
\end{equation}

A plot of the median number of droplets required to cause a spreading event plotted against $(\delta/R)^2$ is shown in FIG.~\ref{fig:avalanches}(g). While there is significant spread in the data and large variation in the number of droplets required to cause a spreading event, the data is consistent with a simple linear fit which agrees with Equation \ref{eqn:avalanche}. This result further demonstrates that the granular capillary length is a fundamental length scale that can be used to describe many properties of our experimental system.

\section{Conclusion}

In conclusion, we have developed an experimental system where we can observe the spreading of a pile of monodisperse oil droplets in two-dimensions. In the system there are both buoyant forces, which act to limit the size of the pile, and adhesive forces, which help stabilize the structure of the pile. Both of these forces can be tuned and we find that both the steady-state structure and the spreading dynamics depend on a balance between these forces. We introduce a granular capillary length, $\delta = \sqrt{\mathscr{A}/{\Delta\rho g \sin{\theta}}}$ in analogy with the capillary length found in liquids. This work expands on our previous work, showing that despite the fact that the oil droplets are granular in nature, with the absence of friction and the presence of an adhesive force, aggregates of these droplets can be described in a manner that would normally be used to describe a continuum liquid. This study enables us to further investigate the boundary between the continuum and the discrete approach for granular systems.

\bibliography{references.bib}

\begin{thebibliography}{46}%
\makeatletter
\providecommand \@ifxundefined [1]{%
 \@ifx{#1\undefined}
}%
\providecommand \@ifnum [1]{%
 \ifnum #1\expandafter \@firstoftwo
 \else \expandafter \@secondoftwo
 \fi
}%
\providecommand \@ifx [1]{%
 \ifx #1\expandafter \@firstoftwo
 \else \expandafter \@secondoftwo
 \fi
}%
\providecommand \natexlab [1]{#1}%
\providecommand \enquote  [1]{``#1''}%
\providecommand \bibnamefont  [1]{#1}%
\providecommand \bibfnamefont [1]{#1}%
\providecommand \citenamefont [1]{#1}%
\providecommand \href@noop [0]{\@secondoftwo}%
\providecommand \href [0]{\begingroup \@sanitize@url \@href}%
\providecommand \@href[1]{\@@startlink{#1}\@@href}%
\providecommand \@@href[1]{\endgroup#1\@@endlink}%
\providecommand \@sanitize@url [0]{\catcode `\\12\catcode `\$12\catcode
  `\&12\catcode `\#12\catcode `\^12\catcode `\_12\catcode `\%12\relax}%
\providecommand \@@startlink[1]{}%
\providecommand \@@endlink[0]{}%
\providecommand \url  [0]{\begingroup\@sanitize@url \@url }%
\providecommand \@url [1]{\endgroup\@href {#1}{\urlprefix }}%
\providecommand \urlprefix  [0]{URL }%
\providecommand \Eprint [0]{\href }%
\providecommand \doibase [0]{https://doi.org/}%
\providecommand \selectlanguage [0]{\@gobble}%
\providecommand \bibinfo  [0]{\@secondoftwo}%
\providecommand \bibfield  [0]{\@secondoftwo}%
\providecommand \translation [1]{[#1]}%
\providecommand \BibitemOpen [0]{}%
\providecommand \bibitemStop [0]{}%
\providecommand \bibitemNoStop [0]{.\EOS\space}%
\providecommand \EOS [0]{\spacefactor3000\relax}%
\providecommand \BibitemShut  [1]{\csname bibitem#1\endcsname}%
\let\auto@bib@innerbib\@empty
\bibitem [{\citenamefont {Andreotti}\ \emph {et~al.}(2013)\citenamefont
  {Andreotti}, \citenamefont {Forterre},\ and\ \citenamefont
  {Pouliquen}}]{BrunoAndreottiYoelForterre2013}%
  \BibitemOpen
  \bibfield  {author} {\bibinfo {author} {\bibfnamefont {B.}~\bibnamefont
  {Andreotti}}, \bibinfo {author} {\bibfnamefont {Y.}~\bibnamefont
  {Forterre}},\ and\ \bibinfo {author} {\bibfnamefont {O.}~\bibnamefont
  {Pouliquen}},\ }\href {https://doi.org/10.1017/CBO9781139541008} {\emph
  {\bibinfo {title} {Granular Media: Between Fluid and Solid}}}\ (\bibinfo
  {publisher} {Cambridge University Press},\ \bibinfo {year}
  {2013})\BibitemShut {NoStop}%
\bibitem [{\citenamefont {Jaeger}\ \emph {et~al.}(1996)\citenamefont {Jaeger},
  \citenamefont {Nagel},\ and\ \citenamefont {Behringer}}]{Jaeger1996}%
  \BibitemOpen
  \bibfield  {author} {\bibinfo {author} {\bibfnamefont {H.~M.}\ \bibnamefont
  {Jaeger}}, \bibinfo {author} {\bibfnamefont {S.~R.}\ \bibnamefont {Nagel}},\
  and\ \bibinfo {author} {\bibfnamefont {R.~P.}\ \bibnamefont {Behringer}},\
  }\href {https://doi.org/10.1103/RevModPhys.68.1259} {\bibfield  {journal}
  {\bibinfo  {journal} {Rev. Mod. Phys.}\ }\textbf {\bibinfo {volume} {68}},\
  \bibinfo {pages} {1259} (\bibinfo {year} {1996})}\BibitemShut {NoStop}%
\bibitem [{\citenamefont {GDR-MiDi}(2004)}]{GDR}%
  \BibitemOpen
  \bibfield  {author} {\bibinfo {author} {\bibnamefont {GDR-MiDi}},\
  }\href@noop {} {\bibfield  {journal} {\bibinfo  {journal} {The European
  Physical Journal E}\ }\textbf {\bibinfo {volume} {14}},\ \bibinfo {pages}
  {341} (\bibinfo {year} {2004})}\BibitemShut {NoStop}%
\bibitem [{\citenamefont {Lemieux}\ and\ \citenamefont
  {Durian}(2000)}]{Lemieux2000}%
  \BibitemOpen
  \bibfield  {author} {\bibinfo {author} {\bibfnamefont {P.~A.}\ \bibnamefont
  {Lemieux}}\ and\ \bibinfo {author} {\bibfnamefont {D.~J.}\ \bibnamefont
  {Durian}},\ }\href {https://doi.org/10.1103/PhysRevLett.85.4273} {\bibfield
  {journal} {\bibinfo  {journal} {Physical Review Letters}\ }\textbf {\bibinfo
  {volume} {85}},\ \bibinfo {pages} {4273} (\bibinfo {year}
  {2000})}\BibitemShut {NoStop}%
\bibitem [{\citenamefont {Frette}\ \emph {et~al.}(1996)\citenamefont {Frette},
  \citenamefont {Christensen}, \citenamefont {Malthe-S{\o}renssen},
  \citenamefont {Feder}, \citenamefont {J{\o}ssang},\ and\ \citenamefont
  {Meakin}}]{Frette1996}%
  \BibitemOpen
  \bibfield  {author} {\bibinfo {author} {\bibfnamefont {V.}~\bibnamefont
  {Frette}}, \bibinfo {author} {\bibfnamefont {K.}~\bibnamefont {Christensen}},
  \bibinfo {author} {\bibfnamefont {A.}~\bibnamefont {Malthe-S{\o}renssen}},
  \bibinfo {author} {\bibfnamefont {J.}~\bibnamefont {Feder}}, \bibinfo
  {author} {\bibfnamefont {T.}~\bibnamefont {J{\o}ssang}},\ and\ \bibinfo
  {author} {\bibfnamefont {P.}~\bibnamefont {Meakin}},\ }\bibfield  {journal}
  {\bibinfo  {journal} {Nature}\ }\href {https://doi.org/10.1038/379049a0}
  {10.1038/379049a0} (\bibinfo {year} {1996})\BibitemShut {NoStop}%
\bibitem [{\citenamefont {Liu}\ \emph {et~al.}(1995)\citenamefont {Liu},
  \citenamefont {Nagel}, \citenamefont {Schecter}, \citenamefont {Coppersmith},
  \citenamefont {Majumdar}, \citenamefont {Narayan},\ and\ \citenamefont
  {Witten}}]{Liu1995}%
  \BibitemOpen
  \bibfield  {author} {\bibinfo {author} {\bibfnamefont {C.~H.}\ \bibnamefont
  {Liu}}, \bibinfo {author} {\bibfnamefont {S.~R.}\ \bibnamefont {Nagel}},
  \bibinfo {author} {\bibfnamefont {D.~A.}\ \bibnamefont {Schecter}}, \bibinfo
  {author} {\bibfnamefont {S.~N.}\ \bibnamefont {Coppersmith}}, \bibinfo
  {author} {\bibfnamefont {S.}~\bibnamefont {Majumdar}}, \bibinfo {author}
  {\bibfnamefont {O.}~\bibnamefont {Narayan}},\ and\ \bibinfo {author}
  {\bibfnamefont {T.~A.}\ \bibnamefont {Witten}},\ }\bibfield  {journal}
  {\bibinfo  {journal} {Science}\ }\href
  {https://doi.org/10.1126/science.269.5223.513} {10.1126/science.269.5223.513}
  (\bibinfo {year} {1995})\BibitemShut {NoStop}%
\bibitem [{\citenamefont {Liu}\ and\ \citenamefont {Nagel}(1998)}]{Liu1998}%
  \BibitemOpen
  \bibfield  {author} {\bibinfo {author} {\bibfnamefont {A.~J.}\ \bibnamefont
  {Liu}}\ and\ \bibinfo {author} {\bibfnamefont {S.~R.}\ \bibnamefont
  {Nagel}},\ }\href@noop {} {\bibfield  {journal} {\bibinfo  {journal}
  {Nature}\ }\textbf {\bibinfo {volume} {396}},\ \bibinfo {pages} {21}
  (\bibinfo {year} {1998})}\BibitemShut {NoStop}%
\bibitem [{\citenamefont {Cates}\ \emph {et~al.}(1998)\citenamefont {Cates},
  \citenamefont {Wittmer}, \citenamefont {Bouchaud},\ and\ \citenamefont
  {Claudin}}]{Cates1998}%
  \BibitemOpen
  \bibfield  {author} {\bibinfo {author} {\bibfnamefont {M.}~\bibnamefont
  {Cates}}, \bibinfo {author} {\bibfnamefont {J.}~\bibnamefont {Wittmer}},
  \bibinfo {author} {\bibfnamefont {J.-P.}\ \bibnamefont {Bouchaud}},\ and\
  \bibinfo {author} {\bibfnamefont {P.}~\bibnamefont {Claudin}},\ }\href@noop
  {} {\bibfield  {journal} {\bibinfo  {journal} {Physical review letters}\
  }\textbf {\bibinfo {volume} {81}},\ \bibinfo {pages} {1841} (\bibinfo {year}
  {1998})}\BibitemShut {NoStop}%
\bibitem [{\citenamefont {Zuriguel}\ \emph {et~al.}(2014)\citenamefont
  {Zuriguel}, \citenamefont {Parisi}, \citenamefont {Hidalgo}, \citenamefont
  {Lozano}, \citenamefont {Janda}, \citenamefont {Gago}, \citenamefont
  {Peralta}, \citenamefont {Ferrer}, \citenamefont {Pugnaloni}, \citenamefont
  {Cl{\'{e}}ment}, \citenamefont {Maza}, \citenamefont {Pagonabarraga},\ and\
  \citenamefont {Garcimart{\'{i}}n}}]{Zuriguel2014}%
  \BibitemOpen
  \bibfield  {author} {\bibinfo {author} {\bibfnamefont {I.}~\bibnamefont
  {Zuriguel}}, \bibinfo {author} {\bibfnamefont {D.~R.}\ \bibnamefont
  {Parisi}}, \bibinfo {author} {\bibfnamefont {R.~C.}\ \bibnamefont {Hidalgo}},
  \bibinfo {author} {\bibfnamefont {C.}~\bibnamefont {Lozano}}, \bibinfo
  {author} {\bibfnamefont {A.}~\bibnamefont {Janda}}, \bibinfo {author}
  {\bibfnamefont {P.~A.}\ \bibnamefont {Gago}}, \bibinfo {author}
  {\bibfnamefont {J.~P.}\ \bibnamefont {Peralta}}, \bibinfo {author}
  {\bibfnamefont {L.~M.}\ \bibnamefont {Ferrer}}, \bibinfo {author}
  {\bibfnamefont {L.~A.}\ \bibnamefont {Pugnaloni}}, \bibinfo {author}
  {\bibfnamefont {E.}~\bibnamefont {Cl{\'{e}}ment}}, \bibinfo {author}
  {\bibfnamefont {D.}~\bibnamefont {Maza}}, \bibinfo {author} {\bibfnamefont
  {I.}~\bibnamefont {Pagonabarraga}},\ and\ \bibinfo {author} {\bibfnamefont
  {A.}~\bibnamefont {Garcimart{\'{i}}n}},\ }\bibfield  {journal} {\bibinfo
  {journal} {Scientific Reports}\ }\href {https://doi.org/10.1038/srep07324}
  {10.1038/srep07324} (\bibinfo {year} {2014})\BibitemShut {NoStop}%
\bibitem [{\citenamefont {Dressaire}\ and\ \citenamefont
  {Sauret}(2017)}]{Dressaire2017}%
  \BibitemOpen
  \bibfield  {author} {\bibinfo {author} {\bibfnamefont {E.}~\bibnamefont
  {Dressaire}}\ and\ \bibinfo {author} {\bibfnamefont {A.}~\bibnamefont
  {Sauret}},\ }\href {https://doi.org/10.1039/c6sm01879c} {\bibfield  {journal}
  {\bibinfo  {journal} {Soft Matter}\ }\textbf {\bibinfo {volume} {13}},\
  \bibinfo {pages} {37} (\bibinfo {year} {2017})}\BibitemShut {NoStop}%
\bibitem [{\citenamefont {Behringer}\ and\ \citenamefont
  {Chakraborty}(2018)}]{Behringer_2018}%
  \BibitemOpen
  \bibfield  {author} {\bibinfo {author} {\bibfnamefont {R.~P.}\ \bibnamefont
  {Behringer}}\ and\ \bibinfo {author} {\bibfnamefont {B.}~\bibnamefont
  {Chakraborty}},\ }\href {https://doi.org/10.1088/1361-6633/aadc3c} {\bibfield
   {journal} {\bibinfo  {journal} {Reports on Progress in Physics}\ }\textbf
  {\bibinfo {volume} {82}},\ \bibinfo {pages} {012601} (\bibinfo {year}
  {2018})}\BibitemShut {NoStop}%
\bibitem [{\citenamefont {Mueth}\ \emph {et~al.}(1998)\citenamefont {Mueth},
  \citenamefont {Jaeger},\ and\ \citenamefont {Nagel}}]{Mueth1998}%
  \BibitemOpen
  \bibfield  {author} {\bibinfo {author} {\bibfnamefont {D.~M.}\ \bibnamefont
  {Mueth}}, \bibinfo {author} {\bibfnamefont {H.~M.}\ \bibnamefont {Jaeger}},\
  and\ \bibinfo {author} {\bibfnamefont {S.~R.}\ \bibnamefont {Nagel}},\
  }\href@noop {} {\bibfield  {journal} {\bibinfo  {journal} {Physical Review
  E}\ }\textbf {\bibinfo {volume} {57}},\ \bibinfo {pages} {3164} (\bibinfo
  {year} {1998})}\BibitemShut {NoStop}%
\bibitem [{\citenamefont {Vallejo}\ \emph {et~al.}(2005)\citenamefont
  {Vallejo}, \citenamefont {Lobo-Guerrero},\ and\ \citenamefont
  {Chik}}]{Vallejo2005}%
  \BibitemOpen
  \bibfield  {author} {\bibinfo {author} {\bibfnamefont {L.~E.}\ \bibnamefont
  {Vallejo}}, \bibinfo {author} {\bibfnamefont {S.}~\bibnamefont
  {Lobo-Guerrero}},\ and\ \bibinfo {author} {\bibfnamefont {Z.}~\bibnamefont
  {Chik}},\ }\href {https://doi.org/10.1007/1-84628-048-6_5} {\emph {\bibinfo
  {title} {Fractals in Engineering: New Trends in Theory and Applications}}}\
  (\bibinfo  {publisher} {Springer-Verlag London},\ \bibinfo {year}
  {2005})\BibitemShut {NoStop}%
\bibitem [{\citenamefont {{Beakawi Al-Hashemi}}\ and\ \citenamefont {{Baghabra
  Al-Amoudi}}(2018)}]{BeakawiAl-Hashemi2018}%
  \BibitemOpen
  \bibfield  {author} {\bibinfo {author} {\bibfnamefont {H.~M.}\ \bibnamefont
  {{Beakawi Al-Hashemi}}}\ and\ \bibinfo {author} {\bibfnamefont {O.~S.}\
  \bibnamefont {{Baghabra Al-Amoudi}}},\ }\href
  {https://doi.org/https://doi.org/10.1016/j.powtec.2018.02.003} {\bibinfo
  {title} {A review on the angle of repose of granular materials}} (\bibinfo
  {year} {2018})\BibitemShut {NoStop}%
\bibitem [{\citenamefont {Lanzerstorfer}(2017)}]{Lanzerstorfer2017}%
  \BibitemOpen
  \bibfield  {author} {\bibinfo {author} {\bibfnamefont {C.}~\bibnamefont
  {Lanzerstorfer}},\ }\href@noop {} {\bibfield  {journal} {\bibinfo  {journal}
  {Granular Matter}\ }\textbf {\bibinfo {volume} {19}},\ \bibinfo {pages} {1}
  (\bibinfo {year} {2017})}\BibitemShut {NoStop}%
\bibitem [{\citenamefont {Sarate}\ \emph {et~al.}(2022)\citenamefont {Sarate},
  \citenamefont {Murthy},\ and\ \citenamefont {Sharma}}]{Sarate2022}%
  \BibitemOpen
  \bibfield  {author} {\bibinfo {author} {\bibfnamefont {P.~S.}\ \bibnamefont
  {Sarate}}, \bibinfo {author} {\bibfnamefont {T.~G.}\ \bibnamefont {Murthy}},\
  and\ \bibinfo {author} {\bibfnamefont {P.}~\bibnamefont {Sharma}},\
  }\href@noop {} {\bibfield  {journal} {\bibinfo  {journal} {Soft Matter}\
  }\textbf {\bibinfo {volume} {18}},\ \bibinfo {pages} {2054} (\bibinfo {year}
  {2022})}\BibitemShut {NoStop}%
\bibitem [{\citenamefont {Fu}\ \emph {et~al.}(2020)\citenamefont {Fu},
  \citenamefont {Chen}, \citenamefont {Ferellec},\ and\ \citenamefont
  {Yang}}]{Fu2020}%
  \BibitemOpen
  \bibfield  {author} {\bibinfo {author} {\bibfnamefont {J.~J.}\ \bibnamefont
  {Fu}}, \bibinfo {author} {\bibfnamefont {C.}~\bibnamefont {Chen}}, \bibinfo
  {author} {\bibfnamefont {J.~F.}\ \bibnamefont {Ferellec}},\ and\ \bibinfo
  {author} {\bibfnamefont {J.}~\bibnamefont {Yang}},\ }\bibfield  {journal}
  {\bibinfo  {journal} {Advances in Civil Engineering}\ }\textbf {\bibinfo
  {volume} {2020}},\ \href {https://doi.org/10.1155/2020/8811063}
  {10.1155/2020/8811063} (\bibinfo {year} {2020})\BibitemShut {NoStop}%
\bibitem [{\citenamefont {Zhou}\ \emph {et~al.}(2002)\citenamefont {Zhou},
  \citenamefont {Xu}, \citenamefont {Yu},\ and\ \citenamefont {Zulli}}]{Zhou}%
  \BibitemOpen
  \bibfield  {author} {\bibinfo {author} {\bibfnamefont {Y.}~\bibnamefont
  {Zhou}}, \bibinfo {author} {\bibfnamefont {B.}~\bibnamefont {Xu}}, \bibinfo
  {author} {\bibfnamefont {A.}~\bibnamefont {Yu}},\ and\ \bibinfo {author}
  {\bibfnamefont {P.}~\bibnamefont {Zulli}},\ }\href
  {https://doi.org/https://doi.org/10.1016/S0032-5910(01)00520-4} {\bibfield
  {journal} {\bibinfo  {journal} {Powder Technology}\ }\textbf {\bibinfo
  {volume} {125}},\ \bibinfo {pages} {45} (\bibinfo {year} {2002})}\BibitemShut
  {NoStop}%
\bibitem [{\citenamefont {Combarros}\ \emph {et~al.}(2014)\citenamefont
  {Combarros}, \citenamefont {Feise}, \citenamefont {Zetzener},\ and\
  \citenamefont {Kwade}}]{Combarros2014}%
  \BibitemOpen
  \bibfield  {author} {\bibinfo {author} {\bibfnamefont {M.}~\bibnamefont
  {Combarros}}, \bibinfo {author} {\bibfnamefont {H.~J.}\ \bibnamefont
  {Feise}}, \bibinfo {author} {\bibfnamefont {H.}~\bibnamefont {Zetzener}},\
  and\ \bibinfo {author} {\bibfnamefont {A.}~\bibnamefont {Kwade}},\ }\href
  {https://doi.org/10.1016/j.partic.2013.04.005} {\bibfield  {journal}
  {\bibinfo  {journal} {Particuology}\ }\textbf {\bibinfo {volume} {12}},\
  \bibinfo {pages} {25} (\bibinfo {year} {2014})}\BibitemShut {NoStop}%
\bibitem [{\citenamefont {Hsiao}\ and\ \citenamefont
  {Pradeep}(2019)}]{Hsiao2019}%
  \BibitemOpen
  \bibfield  {author} {\bibinfo {author} {\bibfnamefont {L.~C.}\ \bibnamefont
  {Hsiao}}\ and\ \bibinfo {author} {\bibfnamefont {S.}~\bibnamefont
  {Pradeep}},\ }\href {https://doi.org/10.1016/J.COCIS.2019.04.003} {\bibfield
  {journal} {\bibinfo  {journal} {Current Opinion in Colloid \& Interface
  Science}\ }\textbf {\bibinfo {volume} {43}},\ \bibinfo {pages} {94} (\bibinfo
  {year} {2019})}\BibitemShut {NoStop}%
\bibitem [{\citenamefont {Hornbaker}\ \emph {et~al.}(1997)\citenamefont
  {Hornbaker}, \citenamefont {Albert}, \citenamefont {Barabási},\ and\
  \citenamefont {Schiffer}}]{Hornbaker1997}%
  \BibitemOpen
  \bibfield  {author} {\bibinfo {author} {\bibfnamefont {D.~J.}\ \bibnamefont
  {Hornbaker}}, \bibinfo {author} {\bibfnamefont {R.}~\bibnamefont {Albert}},
  \bibinfo {author} {\bibfnamefont {I.~A. A.-L.}\ \bibnamefont {Barabási}},\
  and\ \bibinfo {author} {\bibfnamefont {P.}~\bibnamefont {Schiffer}},\
  }\href@noop {} {\bibfield  {journal} {\bibinfo  {journal} {Pilpel, N.
  Manufact. Chem. Aerosol News}\ }\textbf {\bibinfo {volume} {387}},\ \bibinfo
  {pages} {379} (\bibinfo {year} {1997})}\BibitemShut {NoStop}%
\bibitem [{\citenamefont {Gans}\ \emph {et~al.}(2020)\citenamefont {Gans},
  \citenamefont {Pouliquen},\ and\ \citenamefont {Nicolas}}]{Gans2020}%
  \BibitemOpen
  \bibfield  {author} {\bibinfo {author} {\bibfnamefont {A.}~\bibnamefont
  {Gans}}, \bibinfo {author} {\bibfnamefont {O.}~\bibnamefont {Pouliquen}},\
  and\ \bibinfo {author} {\bibfnamefont {M.}~\bibnamefont {Nicolas}},\ }\href
  {https://doi.org/10.1103/PhysRevE.101.032904} {\bibfield  {journal} {\bibinfo
   {journal} {Phys. Rev. E}\ }\textbf {\bibinfo {volume} {101}},\ \bibinfo
  {pages} {032904} (\bibinfo {year} {2020})}\BibitemShut {NoStop}%
\bibitem [{\citenamefont {Halsey}\ and\ \citenamefont
  {Levine}(1998)}]{Halsey1998}%
  \BibitemOpen
  \bibfield  {author} {\bibinfo {author} {\bibfnamefont {T.~C.}\ \bibnamefont
  {Halsey}}\ and\ \bibinfo {author} {\bibfnamefont {A.~J.}\ \bibnamefont
  {Levine}},\ }\href {https://doi.org/10.1103/PhysRevLett.80.3141} {\bibfield
  {journal} {\bibinfo  {journal} {Physical Review Letters}\ }\textbf {\bibinfo
  {volume} {80}},\ \bibinfo {pages} {3141} (\bibinfo {year}
  {1998})}\BibitemShut {NoStop}%
\bibitem [{\citenamefont {Nowak}\ \emph {et~al.}(2005)\citenamefont {Nowak},
  \citenamefont {Samadani},\ and\ \citenamefont {Kudrolli}}]{Nowak2005}%
  \BibitemOpen
  \bibfield  {author} {\bibinfo {author} {\bibfnamefont {S.}~\bibnamefont
  {Nowak}}, \bibinfo {author} {\bibfnamefont {A.}~\bibnamefont {Samadani}},\
  and\ \bibinfo {author} {\bibfnamefont {A.}~\bibnamefont {Kudrolli}},\
  }\href@noop {} {\bibfield  {journal} {\bibinfo  {journal} {Nature Physics}\
  }\textbf {\bibinfo {volume} {1}},\ \bibinfo {pages} {50} (\bibinfo {year}
  {2005})}\BibitemShut {NoStop}%
\bibitem [{\citenamefont {Richefeu}\ \emph {et~al.}(2006)\citenamefont
  {Richefeu}, \citenamefont {Youssoufi},\ and\ \citenamefont
  {Radjaï}}]{Richefeu2006}%
  \BibitemOpen
  \bibfield  {author} {\bibinfo {author} {\bibfnamefont {V.}~\bibnamefont
  {Richefeu}}, \bibinfo {author} {\bibfnamefont {M.~S.~E.}\ \bibnamefont
  {Youssoufi}},\ and\ \bibinfo {author} {\bibfnamefont {F.}~\bibnamefont
  {Radjaï}},\ }\href
  {https://doi.org/10.1103/PHYSREVE.73.051304/FIGURES/15/MEDIUM} {\bibfield
  {journal} {\bibinfo  {journal} {Physical Review E - Statistical, Nonlinear,
  and Soft Matter Physics}\ }\textbf {\bibinfo {volume} {73}},\ \bibinfo
  {pages} {051304} (\bibinfo {year} {2006})}\BibitemShut {NoStop}%
\bibitem [{\citenamefont {Gaudel}\ \emph {et~al.}(2016)\citenamefont {Gaudel},
  \citenamefont {Kiesgen~de Richter}, \citenamefont {Louvet}, \citenamefont
  {Jenny},\ and\ \citenamefont {Skali-Lami}}]{Naima2016}%
  \BibitemOpen
  \bibfield  {author} {\bibinfo {author} {\bibfnamefont {N.}~\bibnamefont
  {Gaudel}}, \bibinfo {author} {\bibfnamefont {S.}~\bibnamefont {Kiesgen~de
  Richter}}, \bibinfo {author} {\bibfnamefont {N.}~\bibnamefont {Louvet}},
  \bibinfo {author} {\bibfnamefont {M.}~\bibnamefont {Jenny}},\ and\ \bibinfo
  {author} {\bibfnamefont {S.}~\bibnamefont {Skali-Lami}},\ }\href
  {https://doi.org/10.1103/PhysRevE.94.032904} {\bibfield  {journal} {\bibinfo
  {journal} {Phys. Rev. E}\ }\textbf {\bibinfo {volume} {94}},\ \bibinfo
  {pages} {032904} (\bibinfo {year} {2016})}\BibitemShut {NoStop}%
\bibitem [{\citenamefont {B\'erut}\ \emph {et~al.}(2019)\citenamefont
  {B\'erut}, \citenamefont {Pouliquen},\ and\ \citenamefont
  {Forterre}}]{Berut2019}%
  \BibitemOpen
  \bibfield  {author} {\bibinfo {author} {\bibfnamefont {A.}~\bibnamefont
  {B\'erut}}, \bibinfo {author} {\bibfnamefont {O.}~\bibnamefont {Pouliquen}},\
  and\ \bibinfo {author} {\bibfnamefont {Y.}~\bibnamefont {Forterre}},\ }\href
  {https://doi.org/10.1103/PhysRevLett.123.248005} {\bibfield  {journal}
  {\bibinfo  {journal} {Phys. Rev. Lett.}\ }\textbf {\bibinfo {volume} {123}},\
  \bibinfo {pages} {248005} (\bibinfo {year} {2019})}\BibitemShut {NoStop}%
\bibitem [{\citenamefont {{Robert De Saint Vincent}}\ \emph
  {et~al.}(2016)\citenamefont {{Robert De Saint Vincent}}, \citenamefont
  {Abkarian},\ and\ \citenamefont {Tabuteau}}]{RobertDeSaintVincent2016}%
  \BibitemOpen
  \bibfield  {author} {\bibinfo {author} {\bibfnamefont {M.}~\bibnamefont
  {{Robert De Saint Vincent}}}, \bibinfo {author} {\bibfnamefont
  {M.}~\bibnamefont {Abkarian}},\ and\ \bibinfo {author} {\bibfnamefont
  {H.}~\bibnamefont {Tabuteau}},\ }\href {https://doi.org/10.1039/c5sm01952d}
  {\bibfield  {journal} {\bibinfo  {journal} {Soft Matter}\ }\textbf {\bibinfo
  {volume} {12}},\ \bibinfo {pages} {1041} (\bibinfo {year}
  {2016})}\BibitemShut {NoStop}%
\bibitem [{\citenamefont {S\'anchez}\ \emph {et~al.}(2007)\citenamefont
  {S\'anchez}, \citenamefont {Raynaud}, \citenamefont {Lanuza}, \citenamefont
  {Andreotti}, \citenamefont {Cl\'ement},\ and\ \citenamefont
  {Aranson}}]{Sanchez2007}%
  \BibitemOpen
  \bibfield  {author} {\bibinfo {author} {\bibfnamefont {I.}~\bibnamefont
  {S\'anchez}}, \bibinfo {author} {\bibfnamefont {F.}~\bibnamefont {Raynaud}},
  \bibinfo {author} {\bibfnamefont {J.}~\bibnamefont {Lanuza}}, \bibinfo
  {author} {\bibfnamefont {B.}~\bibnamefont {Andreotti}}, \bibinfo {author}
  {\bibfnamefont {E.}~\bibnamefont {Cl\'ement}},\ and\ \bibinfo {author}
  {\bibfnamefont {I.~S.}\ \bibnamefont {Aranson}},\ }\href
  {https://doi.org/10.1103/PhysRevE.76.060301} {\bibfield  {journal} {\bibinfo
  {journal} {Phys. Rev. E}\ }\textbf {\bibinfo {volume} {76}},\ \bibinfo
  {pages} {060301} (\bibinfo {year} {2007})}\BibitemShut {NoStop}%
\bibitem [{\citenamefont {Mar\'{\i}n}\ \emph {et~al.}(2011)\citenamefont
  {Mar\'{\i}n}, \citenamefont {Gelderblom}, \citenamefont {Lohse},\ and\
  \citenamefont {Snoeijer}}]{Alvaro2011}%
  \BibitemOpen
  \bibfield  {author} {\bibinfo {author} {\bibfnamefont {A.~G.}\ \bibnamefont
  {Mar\'{\i}n}}, \bibinfo {author} {\bibfnamefont {H.}~\bibnamefont
  {Gelderblom}}, \bibinfo {author} {\bibfnamefont {D.}~\bibnamefont {Lohse}},\
  and\ \bibinfo {author} {\bibfnamefont {J.~H.}\ \bibnamefont {Snoeijer}},\
  }\href {https://doi.org/10.1103/PhysRevLett.107.085502} {\bibfield  {journal}
  {\bibinfo  {journal} {Phys. Rev. Lett.}\ }\textbf {\bibinfo {volume} {107}},\
  \bibinfo {pages} {085502} (\bibinfo {year} {2011})}\BibitemShut {NoStop}%
\bibitem [{\citenamefont {Ortiz}\ \emph {et~al.}(2013)\citenamefont {Ortiz},
  \citenamefont {Riehn},\ and\ \citenamefont {Daniels}}]{Ortiz2013}%
  \BibitemOpen
  \bibfield  {author} {\bibinfo {author} {\bibfnamefont {C.~P.}\ \bibnamefont
  {Ortiz}}, \bibinfo {author} {\bibfnamefont {R.}~\bibnamefont {Riehn}},\ and\
  \bibinfo {author} {\bibfnamefont {K.~E.}\ \bibnamefont {Daniels}},\ }\href
  {https://doi.org/10.1039/c2sm26762d} {\bibfield  {journal} {\bibinfo
  {journal} {Soft Matter}\ }\textbf {\bibinfo {volume} {9}},\ \bibinfo {pages}
  {543} (\bibinfo {year} {2013})}\BibitemShut {NoStop}%
\bibitem [{\citenamefont {Shorts}\ and\ \citenamefont
  {Feitosa}(2018)}]{Shorts2018}%
  \BibitemOpen
  \bibfield  {author} {\bibinfo {author} {\bibfnamefont {D.~C.}\ \bibnamefont
  {Shorts}}\ and\ \bibinfo {author} {\bibfnamefont {K.}~\bibnamefont
  {Feitosa}},\ }\href {https://doi.org/10.1007/s10035-017-0774-x} {\bibfield
  {journal} {\bibinfo  {journal} {Granular Matter}\ }\textbf {\bibinfo {volume}
  {20}},\ \bibinfo {pages} {1} (\bibinfo {year} {2018})}\BibitemShut {NoStop}%
\bibitem [{\citenamefont {Lespiat}\ \emph {et~al.}(2011)\citenamefont
  {Lespiat}, \citenamefont {Cohen-Addad},\ and\ \citenamefont
  {H\"ohler}}]{Lespiat}%
  \BibitemOpen
  \bibfield  {author} {\bibinfo {author} {\bibfnamefont {R.}~\bibnamefont
  {Lespiat}}, \bibinfo {author} {\bibfnamefont {S.}~\bibnamefont
  {Cohen-Addad}},\ and\ \bibinfo {author} {\bibfnamefont {R.}~\bibnamefont
  {H\"ohler}},\ }\href {https://doi.org/10.1103/PhysRevLett.106.148302}
  {\bibfield  {journal} {\bibinfo  {journal} {Phys. Rev. Lett.}\ }\textbf
  {\bibinfo {volume} {106}},\ \bibinfo {pages} {148302} (\bibinfo {year}
  {2011})}\BibitemShut {NoStop}%
\bibitem [{\citenamefont {Perrin}\ \emph {et~al.}(2019)\citenamefont {Perrin},
  \citenamefont {Clavaud}, \citenamefont {Wyart}, \citenamefont {Metzger},\
  and\ \citenamefont {Forterre}}]{Perrin2019}%
  \BibitemOpen
  \bibfield  {author} {\bibinfo {author} {\bibfnamefont {H.}~\bibnamefont
  {Perrin}}, \bibinfo {author} {\bibfnamefont {C.}~\bibnamefont {Clavaud}},
  \bibinfo {author} {\bibfnamefont {M.}~\bibnamefont {Wyart}}, \bibinfo
  {author} {\bibfnamefont {B.}~\bibnamefont {Metzger}},\ and\ \bibinfo {author}
  {\bibfnamefont {Y.}~\bibnamefont {Forterre}},\ }\bibfield  {journal}
  {\bibinfo  {journal} {Physical Review X}\ }\textbf {\bibinfo {volume} {9}},\
  \href {https://doi.org/10.1103/PhysRevX.9.031027} {10.1103/PhysRevX.9.031027}
  (\bibinfo {year} {2019})\BibitemShut {NoStop}%
\bibitem [{\citenamefont {Perrin}\ \emph {et~al.}(2021)\citenamefont {Perrin},
  \citenamefont {Wyart}, \citenamefont {Metzger},\ and\ \citenamefont
  {Forterre}}]{Perrin2021}%
  \BibitemOpen
  \bibfield  {author} {\bibinfo {author} {\bibfnamefont {H.}~\bibnamefont
  {Perrin}}, \bibinfo {author} {\bibfnamefont {M.}~\bibnamefont {Wyart}},
  \bibinfo {author} {\bibfnamefont {B.}~\bibnamefont {Metzger}},\ and\ \bibinfo
  {author} {\bibfnamefont {Y.}~\bibnamefont {Forterre}},\ }\href
  {https://doi.org/10.1103/PhysRevLett.126.228002} {\bibfield  {journal}
  {\bibinfo  {journal} {Phys. Rev. Lett.}\ }\textbf {\bibinfo {volume} {126}},\
  \bibinfo {pages} {228002} (\bibinfo {year} {2021})}\BibitemShut {NoStop}%
\bibitem [{\citenamefont {Peyneau}\ and\ \citenamefont
  {Roux}(2008)}]{Peyneau2008}%
  \BibitemOpen
  \bibfield  {author} {\bibinfo {author} {\bibfnamefont {P.~E.}\ \bibnamefont
  {Peyneau}}\ and\ \bibinfo {author} {\bibfnamefont {J.~N.}\ \bibnamefont
  {Roux}},\ }\bibfield  {journal} {\bibinfo  {journal} {Physical Review E -
  Statistical, Nonlinear, and Soft Matter Physics}\ }\href
  {https://doi.org/10.1103/PhysRevE.78.011307} {10.1103/PhysRevE.78.011307}
  (\bibinfo {year} {2008}),\ \Eprint {https://arxiv.org/abs/0802.1502}
  {arXiv:0802.1502} \BibitemShut {NoStop}%
\bibitem [{\citenamefont {de~Gennes}\ \emph {et~al.}(2004)\citenamefont
  {de~Gennes}, \citenamefont {Brochard-Wyart},\ and\ \citenamefont
  {Qu{\'{e}}r{\'{e}}}}]{DeGennes2004}%
  \BibitemOpen
  \bibfield  {author} {\bibinfo {author} {\bibfnamefont {P.-G.}\ \bibnamefont
  {de~Gennes}}, \bibinfo {author} {\bibfnamefont {F.}~\bibnamefont
  {Brochard-Wyart}},\ and\ \bibinfo {author} {\bibfnamefont {D.}~\bibnamefont
  {Qu{\'{e}}r{\'{e}}}},\ }\href {https://doi.org/10.1007/978-0-387-21656-0}
  {\emph {\bibinfo {title} {Capillarity and Wetting Phenomena}}}\ (\bibinfo
  {publisher} {Springer, New York, NY},\ \bibinfo {year} {2004})\BibitemShut
  {NoStop}%
\bibitem [{\citenamefont {Ono-dit Biot}\ \emph {et~al.}(2020)\citenamefont
  {Ono-dit Biot}, \citenamefont {Lorand},\ and\ \citenamefont
  {Dalnoki-Veress}}]{Ono-dit-Biot2020}%
  \BibitemOpen
  \bibfield  {author} {\bibinfo {author} {\bibfnamefont {J.~C.}\ \bibnamefont
  {Ono-dit Biot}}, \bibinfo {author} {\bibfnamefont {T.}~\bibnamefont
  {Lorand}},\ and\ \bibinfo {author} {\bibfnamefont {K.}~\bibnamefont
  {Dalnoki-Veress}},\ }\href {https://doi.org/10.1103/PhysRevLett.125.228001}
  {\bibfield  {journal} {\bibinfo  {journal} {Physical Review Letters}\
  }\textbf {\bibinfo {volume} {125}},\ \bibinfo {pages} {228001} (\bibinfo
  {year} {2020})}\BibitemShut {NoStop}%
\bibitem [{\citenamefont {Bibette}\ \emph {et~al.}(1992)\citenamefont
  {Bibette}, \citenamefont {Roux},\ and\ \citenamefont
  {Pouligny}}]{Bibette1992}%
  \BibitemOpen
  \bibfield  {author} {\bibinfo {author} {\bibfnamefont {J.}~\bibnamefont
  {Bibette}}, \bibinfo {author} {\bibfnamefont {D.}~\bibnamefont {Roux}},\ and\
  \bibinfo {author} {\bibfnamefont {B.}~\bibnamefont {Pouligny}},\ }\href
  {https://doi.org/10.1051/jp2:1992141} {\bibfield  {journal} {\bibinfo
  {journal} {Journal de Physique II}\ }\textbf {\bibinfo {volume} {2}},\
  \bibinfo {pages} {401} (\bibinfo {year} {1992})}\BibitemShut {NoStop}%
\bibitem [{\citenamefont {Jones}\ \emph {et~al.}(2002)\citenamefont {Jones},
  \citenamefont {Jones}, \citenamefont {Jones} \emph {et~al.}}]{jones2002soft}%
  \BibitemOpen
  \bibfield  {author} {\bibinfo {author} {\bibfnamefont {R.~A.~L.}\
  \bibnamefont {Jones}}, \bibinfo {author} {\bibfnamefont {R.~A.}\ \bibnamefont
  {Jones}}, \bibinfo {author} {\bibfnamefont {R.}~\bibnamefont {Jones}}, \emph
  {et~al.},\ }\href@noop {} {\emph {\bibinfo {title} {Soft condensed
  matter}}},\ Vol.~\bibinfo {volume} {6}\ (\bibinfo  {publisher} {Oxford
  University Press},\ \bibinfo {year} {2002})\BibitemShut {NoStop}%
\bibitem [{\citenamefont {Thevenot}\ \emph {et~al.}(2005)\citenamefont
  {Thevenot}, \citenamefont {Grassl}, \citenamefont {Bastiat},\ and\
  \citenamefont {Binana}}]{Thevenot2005}%
  \BibitemOpen
  \bibfield  {author} {\bibinfo {author} {\bibfnamefont {C.}~\bibnamefont
  {Thevenot}}, \bibinfo {author} {\bibfnamefont {B.}~\bibnamefont {Grassl}},
  \bibinfo {author} {\bibfnamefont {G.}~\bibnamefont {Bastiat}},\ and\ \bibinfo
  {author} {\bibfnamefont {W.}~\bibnamefont {Binana}},\ }\href
  {https://doi.org/https://doi.org/10.1016/j.colsurfa.2004.10.062} {\bibfield
  {journal} {\bibinfo  {journal} {Colloids and Surfaces A: Physicochemical and
  Engineering Aspects}\ }\textbf {\bibinfo {volume} {252}},\ \bibinfo {pages}
  {105} (\bibinfo {year} {2005})}\BibitemShut {NoStop}%
\bibitem [{\citenamefont {Naskar}\ \emph {et~al.}(2013)\citenamefont {Naskar},
  \citenamefont {Dey}, \citenamefont {Satya},\ and\ \citenamefont
  {Moulik}}]{Naskar2013}%
  \BibitemOpen
  \bibfield  {author} {\bibinfo {author} {\bibfnamefont {B.}~\bibnamefont
  {Naskar}}, \bibinfo {author} {\bibfnamefont {A.}~\bibnamefont {Dey}},
  \bibinfo {author} {\bibnamefont {Satya}},\ and\ \bibinfo {author}
  {\bibfnamefont {P.}~\bibnamefont {Moulik}},\ }\href
  {https://doi.org/10.1007/s11743-013-1449-1} {\bibfield  {journal} {\bibinfo
  {journal} {J Surfact Deterg}\ }\textbf {\bibinfo {volume} {16}},\ \bibinfo
  {pages} {785} (\bibinfo {year} {2013})}\BibitemShut {NoStop}%
\bibitem [{\citenamefont {Barkley}\ \emph {et~al.}(2015)\citenamefont
  {Barkley}, \citenamefont {Weeks},\ and\ \citenamefont
  {Dalnoki-Veress}}]{Barkley2015}%
  \BibitemOpen
  \bibfield  {author} {\bibinfo {author} {\bibfnamefont {S.}~\bibnamefont
  {Barkley}}, \bibinfo {author} {\bibfnamefont {E.~R.}\ \bibnamefont {Weeks}},\
  and\ \bibinfo {author} {\bibfnamefont {K.}~\bibnamefont {Dalnoki-Veress}},\
  }\href {https://doi.org/10.1140/epje/i2015-15138-8} {\bibfield  {journal}
  {\bibinfo  {journal} {European Physical Journal E}\ }\textbf {\bibinfo
  {volume} {38}},\ \bibinfo {pages} {1} (\bibinfo {year} {2015})},\ \Eprint
  {https://arxiv.org/abs/1508.03259} {arXiv:1508.03259} \BibitemShut {NoStop}%
\bibitem [{\citenamefont {Barkley}\ \emph {et~al.}(2016)\citenamefont
  {Barkley}, \citenamefont {Scarfe}, \citenamefont {Weeks},\ and\ \citenamefont
  {Dalnoki-Veress}}]{Barkley2016}%
  \BibitemOpen
  \bibfield  {author} {\bibinfo {author} {\bibfnamefont {S.}~\bibnamefont
  {Barkley}}, \bibinfo {author} {\bibfnamefont {S.~J.}\ \bibnamefont {Scarfe}},
  \bibinfo {author} {\bibfnamefont {E.~R.}\ \bibnamefont {Weeks}},\ and\
  \bibinfo {author} {\bibfnamefont {K.}~\bibnamefont {Dalnoki-Veress}},\ }\href
  {https://doi.org/10.1039/c6sm00853d} {\bibfield  {journal} {\bibinfo
  {journal} {Soft Matter}\ }\textbf {\bibinfo {volume} {12}},\ \bibinfo {pages}
  {7398} (\bibinfo {year} {2016})},\ \Eprint {https://arxiv.org/abs/1604.05646}
  {arXiv:1604.05646} \BibitemShut {NoStop}%
\bibitem [{\citenamefont {Ono-Dit-Biot}\ \emph {et~al.}(2020)\citenamefont
  {Ono-Dit-Biot}, \citenamefont {Soulard}, \citenamefont {Barkley},
  \citenamefont {Weeks}, \citenamefont {Salez}, \citenamefont {Raphael},\ and\
  \citenamefont {Dalnoki-Veress}}]{Ono-dit-Biot2021PRR}%
  \BibitemOpen
  \bibfield  {author} {\bibinfo {author} {\bibfnamefont {J.-C.}\ \bibnamefont
  {Ono-Dit-Biot}}, \bibinfo {author} {\bibfnamefont {P.}~\bibnamefont
  {Soulard}}, \bibinfo {author} {\bibfnamefont {S.}~\bibnamefont {Barkley}},
  \bibinfo {author} {\bibfnamefont {E.~R.}\ \bibnamefont {Weeks}}, \bibinfo
  {author} {\bibfnamefont {T.}~\bibnamefont {Salez}}, \bibinfo {author}
  {\bibfnamefont {E.}~\bibnamefont {Raphael}},\ and\ \bibinfo {author}
  {\bibfnamefont {K.}~\bibnamefont {Dalnoki-Veress}},\ }\href@noop {}
  {\bibfield  {journal} {\bibinfo  {journal} {Physical Review Research}\
  }\textbf {\bibinfo {volume} {2}},\ \bibinfo {pages} {023070} (\bibinfo {year}
  {2020})}\BibitemShut {NoStop}%
\bibitem [{\citenamefont {Ono-dit Biot}\ \emph {et~al.}(2021)\citenamefont
  {Ono-dit Biot}, \citenamefont {Soulard}, \citenamefont {Barkley},
  \citenamefont {Weeks}, \citenamefont {Salez}, \citenamefont {Rapha{\"e}l},\
  and\ \citenamefont {Dalnoki-Veress}}]{Ono-dit-Biot2021SM}%
  \BibitemOpen
  \bibfield  {author} {\bibinfo {author} {\bibfnamefont {J.-C.}\ \bibnamefont
  {Ono-dit Biot}}, \bibinfo {author} {\bibfnamefont {P.}~\bibnamefont
  {Soulard}}, \bibinfo {author} {\bibfnamefont {S.}~\bibnamefont {Barkley}},
  \bibinfo {author} {\bibfnamefont {E.~R.}\ \bibnamefont {Weeks}}, \bibinfo
  {author} {\bibfnamefont {T.}~\bibnamefont {Salez}}, \bibinfo {author}
  {\bibfnamefont {{\'E}.}~\bibnamefont {Rapha{\"e}l}},\ and\ \bibinfo {author}
  {\bibfnamefont {K.}~\bibnamefont {Dalnoki-Veress}},\ }\href@noop {}
  {\bibfield  {journal} {\bibinfo  {journal} {Soft Matter}\ }\textbf {\bibinfo
  {volume} {17}},\ \bibinfo {pages} {1194} (\bibinfo {year}
  {2021})}\BibitemShut {NoStop}%
\end{thebibliography}%

\end{document}